# Analysis of Software Engineering for Agile Machine Learning Projects

Kushal Singla, Joy Bose, Chetan Naik
Samsung R&D Institute
Bangalore, India
kushal.s@samsung.com

*Abstract*— The number of machine learning, artificial intelligence or data science related software engineering projects using Agile methodology is increasing. However, there are very few studies on how such projects work in practice. In this paper, we analyze project issues tracking data taken from Scrum (a popular tool for Agile) for several machine learning projects. We compare this data with corresponding data from non-machine learning projects, in an attempt to analyze how machine learning projects are executed differently from normal software engineering projects. On analysis, we find that machine learning project issues use different kinds of words to describe issues, have higher number of exploratory or research oriented tasks as compared to implementation tasks, and have a higher number of issues in the product backlog after each sprint, denoting that it is more difficult to estimate the duration of machine learning project related tasks in advance. After analyzing this data, we propose a few ways in which Agile machine learning projects can be better logged and executed, given their differences with normal software engineering projects.

## I. INTRODUCTION

The number of software engineering projects having some artificial intelligence (AI), machine learning (ML) or data science (DS) related component is growing. One study forecast that the spending on machine learning projects would grow from $12 billion in 2017 to $58 Billion by 2021 [1].

Agile methodology [2] has been very popular recently as a means for software development, because of advantages such as flexibility and rapid prototyping cycles. It has become the de facto standard for many new projects [3] at many big companies. Each of the requirements or wish list of the software product is tracked as the "Product Backlog" in agile. The requirement is termed as "User Story". In this methodology, pending work is divided into units of stories, tasks and subtasks, to be completed in fixed time units or cycles called sprints. This makes it easier to efficiently track the work progress and get customer feedback.

There have been a few studies so far at the effectiveness of agile methodology for software development, but to our knowledge they do not focus on machine learning software related projects. Since machine learning is still an emerging field with its own best practices and processes, it is useful to examine the effectiveness of agile methodology when applied to machine learning projects as opposed to ordinary software projects.

In this paper, we investigate data gathered from two teams within our organization, one machine learning (ML) related and the other not (Non-ML). Both of the teams follow agile methodology in executing their projects. Our objective is to study the characteristics of machine learning projects executed by following agile methodology. Here we are interested in understanding what are the unique features of agile machine learning projects as compared to non-agile, and how do such projects perform when measured with various parameters. We used data from stories, tasks and subtasks collected from a popular scrum tool, named Jira, for our investigation.

The rest of this paper is organized as follows: in section 2 we review related work in the area of agile methodology and machine learning projects. In section 3 we examine our method for data collection on ML and non ML agile projects. Section 4 describes some results from our data analysis. Section 5 discusses ways in which agile methodology could be made more useful and relevant for machine learning projects. Finally, section 6 concludes the paper.

## II. RELATED WORK

There are a number of online articles and blog posts [4-7] examining the feasibility and issues connected with agile machine learning. Most of them provide advice on how to get a successful machine learning project that takes advantage of agile methodology.

Developers at Microsoft proposed a framework called Team Data Science Process (TDSP) [13] to plan a data science or analytics project efficiently using Agile principles, using facilities to incorporate built in machine learning specific terminology such as feature engineering, data exploration etc. Bullock [14] argued that ML projects can benefit from boosted productivity gained from agile processes.

However, all of the above works lack in studies of real agile software development systems and the issues faced by them.

Schleier-Smith [8] studied the feasibility of agile machine learning in real time apps, taking as example the development and design of a dating application on a mobile phone. The study flagged challenges such as difficulty in getting training data and long deployment cycles. It also proposed a generic agile cycle for machine learning projects including continuously observing problems with existing ML models, defining new features, followed by training, testing and redeploying new ML models with the new features.

However, this too did not collect data on the project lifecycle characteristics of real life agile ML related development projects.

There are also a few studies of machine learning techniques applied to different use cases related to software engineering. The study by Wen et. al. [9] reviewed the use of machine learning models applied to estimation of how long it might take to complete software projects. Malhotra [10] applied machine learning methods for to improve software quality by predicting faults.

There are also a number of empirical studies of agile systems, such as by Dyba [11], but they do not cover machine learning projects as a separate category. This is the gap we seek to fill in this paper.

## III. OUR APPROACH FOR DATA COLLECTION

In order to study issues and characteristics related to machine learning projects using agile methodology, we collected data over a period of 6 months from two kinds of teams, one working on exclusively machine learning projects and consisting mainly of data engineers and data scientists, and the other working on conventional software projects (mainly analytics related) not involving machine learning specifically, and consisting largely of software engineers. For convenience, we call the ML related team as team A and the conventional software (non-ML) team as team B.

Both the teams are of similar size (around 15 members each) and composition (experience range between 1-10 years, age between 22-35 years, consisting of male engineers and one female in the non-machine learning team). Both the teams followed agile methodology using scrum, with daily stand up meetings, two week long sprints, a sprint planning meeting at the beginning of each sprint and a demo and retrospective meeting at the end of the sprint.

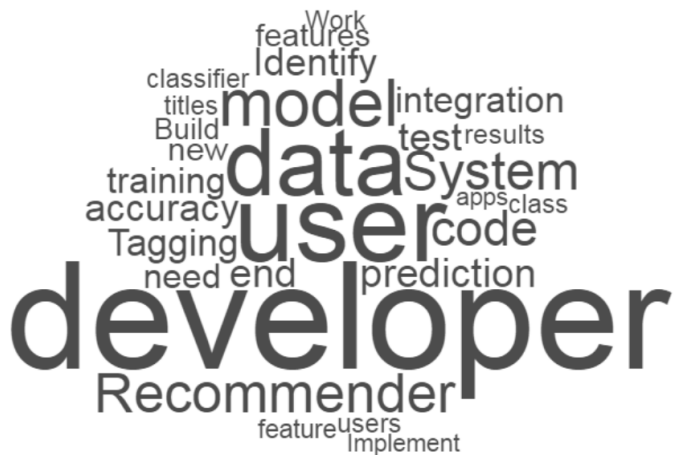

Fig. 1. Word cloud (generated by wordclouds dot com) of words taken from the story summaries from projects of a machine learning team (team A) over a 6 month period

We collected data related to 22 sprint cycles, using the logs from the Jira software [12], a popular issue tracking and product management software developed by Atlassian for Agile projects.

Team A data consisted of 1000 issues including 45 epics, 271 stories and 684 subtasks, and spanning over 8 different projects.

Team B data consisted of 345 issues including 9 epics, 88 stories, 247 sub tasks including bug subtasks, and spanning over 5 projects.

We exported the JIRA logs in Microsoft Excel format, and analyzed them. The team A log came to around 84 MB, and team B log to around 28 MB.

In the following section, we analyze the logs and discuss how we can detect some trends in machine learning projects on the basis of the logs.

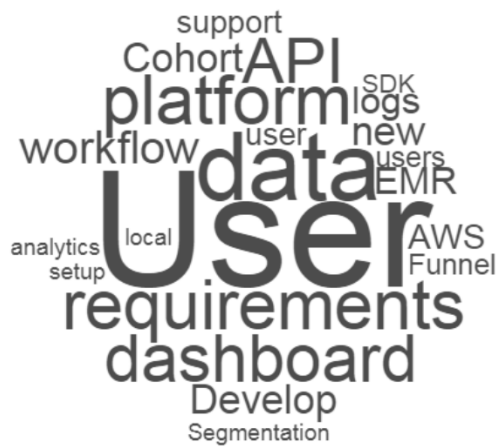

Fig. 2. Word cloud (generated by wordclouds dot com) of words taken from the story summaries from projects of a non machine learning team (team B) over a 6 month period

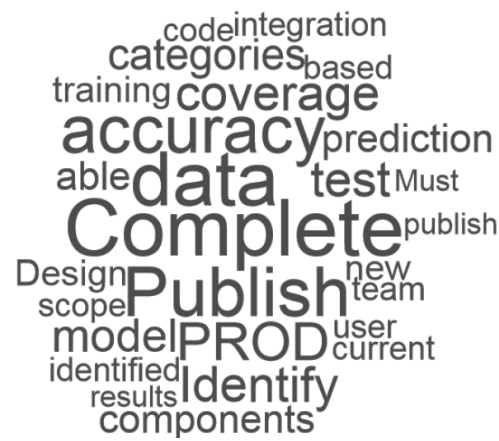

Fig. 3. Word cloud (generated by wordclouds dot com) of words taken from the acceptance criteria of stories from projects of the machine learning team (team A) over a 6 month period.

## IV. RESULTS FROM THE DATA ANALYSIS

In this section, we study the results of our analysis on the collected Jira data related to the software engineering issues and project planning for the two teams, machine learning team A and the non-ML team B.

### A. Analysis of the word clouds generated from the issues

In one study, we generated word clouds from the text of the issue summaries and descriptions of the user stories and subtasks from the Jira tool.

Figs. 1 and 2 show the results of the word clouds generated from the titles of the stories ML team A and the non-ML team B data. Fig. 3 shows the word cloud created from the text of the acceptance criteria from the ML team. The acceptance criteria text for the stories was chosen because it was more likely to be firm and measurable, as opposed to just the title.

As we can see, team A used more machine learning related terms such as tagging, prediction, classifier, accuracy, features, recommender etc. Team B used general software engineering related words such as API, SDK, logs, requirements, AWS, dashboard etc.

From this, we can conclude that machine learning teams may tend to use more ML related terms as opposed to software development terms, when writing their user stories and subtasks.

Also, team A used the word 'developer' most in their stories ('as a developer, I ..') while team B used the term 'user' most, indicating the type of end user or client for whom they intended their software to be used. The machine learning team members generally saw their main role as developers, focusing the software to be consumed internally, while team B was more focused on the external end users.

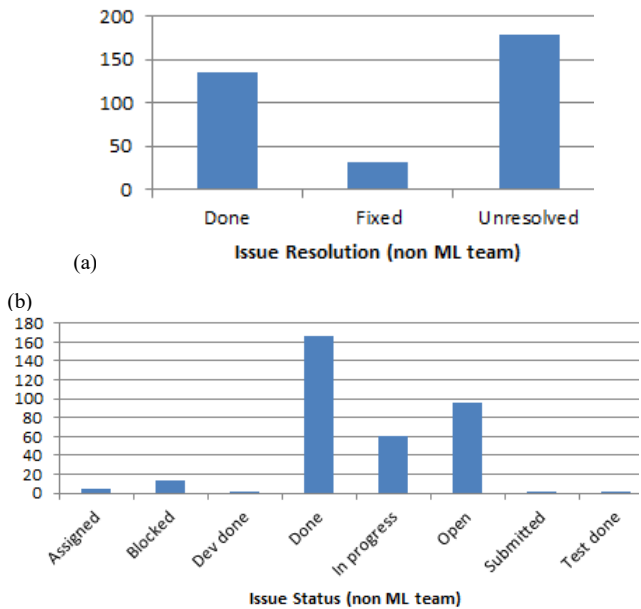

Fig. 4. Plot of the (a) issue resolution and (b) issue status for the non ML team over 22 sprints. The non ML team seemed to have more open and unresolved issues.

### B. Analysis of the backlog after each sprint

We then analyzed the amount of time spent in completion of each story, and how many of the tasks went to the backlog after each sprint.

We found that a much higher number of issues went to the backlog on average for the ML team A as compared to the non ML team B.

From this, we conclude that ML projects may have a higher number of issues in the product backlog after each sprint, denoting that it is more difficult to estimate the duration of machine learning project related tasks in advance. This is because ML project outcomes are not always clear as compared to conventional software projects. A typical ML project may involve applying different machine learning algorithms on the data to make some predictions or recommendations, and depending on the accuracy of the predictions one may need to repeat the machine learning task or experiment with a different algorithm or trying a different approach.

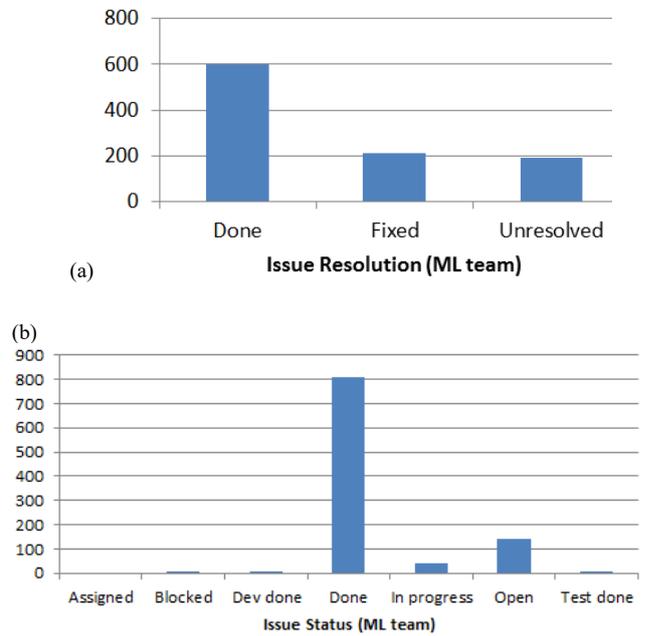

Fig. 5. Plot of the (a) issue resolution and (b) issue status for the ML team over 22 sprints

For example, if the accuracy with a linear SVM is low for a prediction task, one may need to experiment with deep learning model for the same task in order to get a satisfactory prediction accuracy. In ML tasks, the type of the data plays an important role, since one algorithm may be good for a certain type of data, but may give less accuracy if the data is changed. Conversely, conventional software development projects may tend to have more clear-cut outcomes, since the algorithm is well defined and the performance of the algorithm is largely independent of the data.

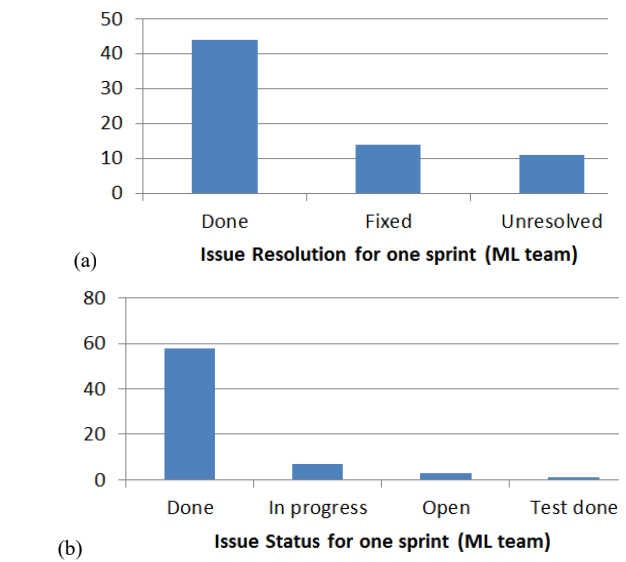

Fig. 6. Plot of the (a) issue resolution and (b) issue status for the ML team over a single sprint. While most issues are resolved as fixed, a few are in the unresolved state as well.

Figures 4-6 give the issue resolution and issue status breakdown for the non ML team and the ML team respectively. The non-ML team had more number of unresolved issues.

*C. Analysis of the average time to complete each story*

We then analyzed the average time spent for each story for ML and non-ML projects. The time spent is taken from the original estimate of the story as shown in the Jira logs. We obtained an average time of 144700 seconds for stories from the ML projects and 84668 seconds for the stories from the non-ML projects. Fig. 7 shows the plot of the same.

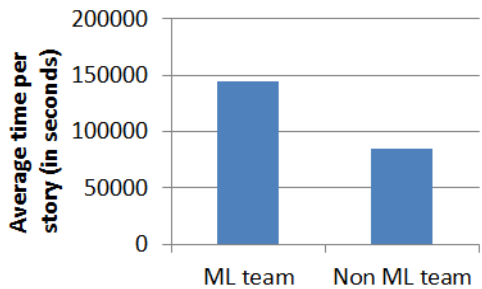

Fig. 7. Plot of the average time to complete each story, in seconds, for the ML projects compared to the non ML projects

As we can see, the average time to complete each story was higher for the ML team A as compared to the non-ML team B. This could be due to more tasks and subtasks ending in the backlog for the ML related team A.

*D. Analysis of the deviation of actual time to complete a story vs the estimated time*

We then computed the average of the deviation from the estimated time to complete each story, for both the ML and the non ML cases. Table 1 shows the average of the original estimate – time actually spent for the ML team and the non ML team. As we can see, the time deviation in actual time from estimated time was higher in case of the ML project.

TABLE I. ANALYSIS OF ORIGINAL TIME ESTIMATE – TIME SPENT PER STORY (IN SECONDS) FOR ML VS NON ML PROJECTS

| Team Type | Mean | Variance | Std. Dev |
|---|---|---|---|
| ML team | 64952.1 | 12865606157.5 | 113426.7 |
| Non ML team | 10813.6 | 866503450.4 | 29436.4 |

*E. Distribution of exploratory (research) vs developmental tasks*

From an analysis of the projects as well as the stories and subtasks within each project, we found that the ML team had more exploratory stories and subtasks as compared to development or testing related tasks.

Such exploratory or research oriented tasks could be identified by verbs such as identify, compare, improve, consider, discuss, study, improve etc.

Development and testing tasks could be identified by verbs such as implement, test, measure, build, integrate, etc.

*F. Shape of the burndown chart*

Fig. 8 gives one screenshot of the burndown chart for the ML team A.

We found that the burndown chart for the ML team was generally more irregular than the non ML team. This further reinforced the observation that ML team task and subtask completion could be more irregular due to the lesser number of clear cut outcomes.

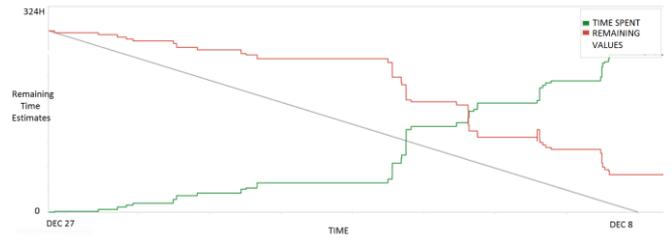

Fig. 8. Burndown chart for a machine learning team A for a single sprint, showing irregularities in the time remaining (in red) and actual time spent (in green) to complete the tasks.

*G. Analysis of the logs from the sprint retrospective meetings*

We found that the ML team had a higher number of words such as overestimation, unexpected work, delayed, ad-hoc work, waiting for, not accounted for, unplanned tasks, more than expected, irregular, etc. in the 'what went wrong' section of the sprint retrospective meetings. The use of such words could further indicate that the ML projects and teams may have less certainty in the tasks due to the more exploratory nature of machine learning related tasks.

*H. Learnings from the analysis: difference between ML and non-ML projects*

From the analysis of the data in the previous subsections, we can see that there are important differences between the implementation of Agile software engineering projects in a machine learning team compared to a non-machine learning team.

At least some of these differences might be explained by the nature of machine learning tasks as compared to non-machine learning tasks. Many machine learning projects cannot predict in advance how well the accuracy of a classification task or regression or relevance of a recommendation will be, since it is dependent on data.

However, some of the differences could also be partly explained by other factors, such as the way the sprint planning was done, the regularity at which issues are logged, the frequency of updating the Jira tool, the accuracy of the data logged, etc.

V. SUGGESTIONS TO MAKE AGILE METHODOLOGY MORE USEFUL OR RELEVANT FOR ML PROJECTS

In this section, based on the previous analysis, we propose a few suggestions in which agile methodology can be made more useful for machine learning projects and teams.

Some of our recommendations are as follows:

- Having a separate category, with separate logging, for exploratory (research oriented) tasks as opposed to development or testing tasks

- Having a best practice of logging of actual status (as opposed to expected status) of each subtask or story

- Having a best practice of updating the status logs on a daily basis before each standup meeting, or as soon as the task is actually completed, instead of waiting for the weekend or end of the sprint.

- Having a best practice of framing the subtask or story descriptions, even the exploratory ones, in more SMART choice of words (such as measure, determine) instead of using vague words such as explore.

## VI. Conclusion

In this paper, we have studied the logs related to software engineering following the agile methodology for a machine learning team and compared it with the logs for a non-machine learning team, analyzing the trends and their reasons. We have also provided a few suggestions by which Agile can be made more useful or relevant for machine learning teams and projects.

In future, we plan to have a bigger study with more teams to get a more generalized insight on the trends. We also plan to study the impact of our recommendations by studying the Jira logs with and without following these recommendations.